\def   \ni {\noindent}

\def   \ssk {\vskip  5truept}

\def   \bsk {\vskip 15truept}

\def   \newline {\hfil\break}

\def\lesssim{\mathrel{\hbox{\rlap{\hbox{\lower4pt\hbox{$\sim$}}}\hbox{$<$}}}}
\def\gtrsim{\mathrel{\hbox{\rlap{\hbox{\lower4pt\hbox{$\sim$}}}\hbox{$>$}}}}

\def\sun{\hbox{$\odot$}}
\def\ion#1#2{#1$\;${\small\rm I}\relax}
\def\halpha{${\rm H} \alpha$}

\def\RXTE{{\it RXTE}}
\def\CGRO{{\it CGRO}}

\documentstyle[epsfig]{article}
\begin{document}

\hsize 5truein
\vsize 8truein
\font\abstract=cmr8
\font\keywords=cmr8
\font\caption=cmr8
\font\references=cmr8
\font\text=cmr10
\font\affiliation=cmssi10
\font\author=cmss10
\font\mc=cmss8
\font\title=cmssbx10 scaled\magstep2
\font\alcit=cmti7 scaled\magstephalf
\font\alcin=cmr6 
\font\ita=cmti8
\font\mma=cmr8
\def\ref{\par\noindent\hangindent 15pt}


\title{\ni OBSERVATIONS OF GX 339--4 IN 1996}
\bsk \bsk
\author{\ni I. A. Smith$^{1}$, E. P. Liang$^{1}$, D. Lin$^{1}$,
M. B\"ottcher$^{1}$, M. Moss$^{1,2}$, A. Crider$^{1}$, A. V. Filippenko$^{3}$,
D. C. Leonard$^{3}$, R. P. Fender$^{4}$, Ph. Durouchoux$^{5}$,
S. Corbel$^{5}$, R. Sood$^{6}$}
\bsk
\affiliation{1) Department of Space Physics and Astronomy, 
Rice University, MS 108, 6100 South Main Street, Houston, TX 77005-1892, USA}

\affiliation{2) Currently at McDonnell-Douglas, St. Louis, MO, USA}

\affiliation{3) Dept. of Astronomy, University of California, Berkeley, 
CA 94720-3411, USA}

\affiliation{4) Astronomical Institute `Anton Pannekoek', Center for 
High-Energy Astrophysics, University of Amsterdam, Kruislaan 403, 
1098 SJ Amsterdam, The Netherlands}

\affiliation{5) DAPNIA, Service d'Astrophysique, CE Saclay,
91191 Gif sur Yvette Cedex, France}

\affiliation{6) School of Physics, ADFA, Northcott Drive, Canberra ACT 2600,
Australia}
\bsk
\baselineskip = 12pt

\abstract{ABSTRACT \ni
GX 339--4 is an unusual black hole candidate in that it is a persistent 
source, being detected by X-ray telescopes most of the time, but it also 
has nova-like flaring states.
In 1996 we performed radio, optical, X-ray, and gamma-ray observations of 
GX 339--4 when it was in a hard state (= soft X-ray low state).
Here we present a brief summary of some of the results.
} 
\bsk
\baselineskip = 12pt
\keywords{\ni KEYWORDS: 
binaries: spectroscopic --- stars individual (GX 339--4) ---
black hole physics --- X-rays: stars --- accretion: accretion disks}

\bsk
\baselineskip = 12pt


\text{\ni 1. KECK OBSERVATIONS
\ssk
\ni     
We obtained optical spectra using the Low-Resolution Imaging Spectrometer 
at the Cassegrain focus of the Keck I telescope on 1996 May 12 UT (MJD 50215).
Full details are given in Smith, Filippenko, \& Leonard (1999).

\begin{figure}
\end{figure}


At this time, neither the ASM on the {\it Rossi X-Ray Timing Explorer} 
(\RXTE) nor BATSE on the {\it Compton Gamma-Ray Observatory} (\CGRO)
detected the source.
We had hoped that the lack of detectable high-energy emission would make 
it possible to observe the companion star for the first time.
However, we found that the optical emission was still dominated by the
accretion disk with $V \approx 17$ mag.
This is typical for the source in the low state, ${\rm V} \approx 15.4 - 20.2$
(Tanaka \& Lewin 1995).

The dominant emission line in our spectra is \halpha.
Its equivalent width (EW) of $\sim 6.5$ \AA\ is similar to the lines seen
in previous GX 339--4 observations 
(Grindlay 1979, Cowley, Crampton, \& Hutchings 1987).
However, this is small compared to other black hole X-ray novae (BHXRN); 
for example, the \halpha\ line in Nova Oph 1977 had EW $= 85$ \AA\ during 
our observing run (Filippenko et al. 1997).

For the first time in GX 339--4 we were able to resolve the \halpha\ line 
and show it has a double peaked profile.
The peak separation implies a $\Delta v = 370 \pm 40 ~{\rm km~s}^{-1}$.
If we assume this peak emission comes from a circular Keplerian orbit,
it would be at a distance $4 \times 10^{11} (M/M_{\sun})$ cm, 
where $M$ is the mass of the black hole.

Double peaked \halpha\ emission lines have been seen in the quiescent 
optical counterparts of many black hole X-ray novae.
The peak separations in these sources are all 
surprisingly similar ($\sim 900 - 1400 ~{\rm km~s}^{-1}$).
The \halpha\ peak separation in GX 339--4 is therefore quite 
different from in the usual BHXRN.
The narrower \halpha\ peak separation in GX 339--4 implies that the
optical emission comes from a larger Keplerian radius than in the BHXRN, 
which may be a clue to its different behavior.

GX 339--4 is most often compared to Cygnus X-1.
Unfortunately, for Cygnus X-1 the absorption and emission lines from the 
companion star dominate, and only the wings of the \halpha\ emission line are 
detected (Canalizo et al. 1995).
However, the width of the base of the \halpha\ emission line in Cygnus X-1
is quite similar to that in GX 339--4, suggesting that the optical
emission from their accretion disks is similar.
It is also interesting to note that the peak separation was
$\sim 350 - 550 ~{\rm km~s}^{-1}$ in the \halpha\ emission lines
from GRO J1655--40 {\it during outbursts} (Soria et al. 1998).

The only other emission lines in our spectra are from \ion{He}{1}.
For $\lambda$5875 and $\lambda$6678 the equivalent widths are
$1.3 \pm 0.3$ \AA\ and $1.0 \pm 0.1$ \AA\ respectively.
We do not see any evidence for an absorption line from 
\ion{Li}{1} at $\lambda$6708 \AA.

\bsk
\ni 2. RADIO OBSERVATIONS
\ssk
\ni     
High resolution 3.5 cm observations of the counterpart using the 
Australia Telescope Compact Array detected a possible jet-like feature 
1996 July 11--13 (MJD 50275--7; Fender et al. 1997).
The radio spectrum is approximately flat, and shows a significant variability
(Smith et al. 1999).
The spectral shape and amplitude were not anomalous for this
source during the time of the possible radio jet.

\bsk
\ni 3. OSSE OBSERVATIONS
\ssk
\ni     
Our OSSE pointed observation was made 1996 July 9-23 (MJD 50273--287).
Full details are given in Smith et al. (1999).
GX 339--4 was in a hard state (= soft X-ray low state).
Unlike in 1996 May, the RXTE ASM and BATSE detected the source, and the OSSE
flux was $\sim 150$ mCrab between 50 and 400 keV.

The daily light curves showed that the OSSE fluxes in the 50--70 and 
70--270 keV bands were generally rising;
the X-ray and gamma-ray fluxes peaked on $\sim$ TJD 50290
(Smith et al. 1997, Rubin et al. 1998).
The spectrum may be softening during the two week 
observation, but the result is not statistically significant.
There was no significant change in the X- and gamma-ray
flux or hardness during the time the possible radio jet-like feature was seen.

The OSSE spectrum was extracted by averaging over the whole two week
observation.
Since the hardness ratio did not change significantly over this time,
this gives a reliable measure of the spectral shape during these
two weeks.
A power law times exponential (PLE) function gave a good fit to the OSSE data.
In contrast to Grabelsky et al. (1995), a Sunyaev-Titarchuk function 
(ST; Sunyaev \& Titarchuk 1980) gave an equally good fit to the OSSE 
data alone.

\bsk
\ni 4. RXTE SPECTROSCOPY
\ssk
\ni     
Our \RXTE\ pointed observation was made 
on 1996 July 26 (MJD 50290), just after the OSSE run ended.
We generated the \RXTE\ spectrum using both the Proportional 
Counter Array (PCA), and the High Energy X-ray Timing Experiment (HEXTE).
Although the source was extremely variable during our observation, the
hardness ratios suggested there was little change in the spectral shape,
so we combined all the data to make one spectrum.
Full details are given in Smith et al. (1999).

A PLE model fits the \RXTE\ data (alone) above 15 keV.
Two extra components are required to fit the spectrum at lower energies:

\begin{itemize}

\item 
A soft component that peaks at $\sim 2.5$ keV, whose exact form is poorly 
determined because the PCA response matrix is unreliable below 2 keV.

\item
A {\it broad} emission centered on $\sim 6.4$ keV.
This may be an iron line that is broadened by orbital Doppler motions and/or
scattering off a hot medium.
Its equivalent width is $\sim 600$ eV.

\end{itemize}

A ST model also gives a good fit to the \RXTE\ data {\it alone}.
However, normalizing and combining the \RXTE\ and OSSE data we found that 
the ST model drops off too rapidly with increasing energies 
to give an acceptable joint fit.

The PLE best fits to the separate OSSE and \RXTE\ data used the same 
$kT$ and had consistent values of $\alpha$.
We therefore get a good joint \RXTE--OSSE fit using the PLE model.
Unlike for the {\it Ginga}--OSSE observation in 1991 (Zdziarski et al. 1998)
our simplistic continuum fitting does not require a
significant reflection component.
This is similar to the result of Dove et al. (1998) who found that no 
reflection component was needed to explain their \RXTE\ observations of 
Cygnus X-1.

Using our detailed self-consistent accretion disk corona models,
we similarly conclude that fitting the GX 339--4 spectral data
does not require a strong reflection component (B\"ottcher et al. 1998).
This is because most of the incident flux from the corona goes into 
heating of the disk surface layer and is not reflected.

\bsk
\ni 5. RXTE RAPID VARIABILITY
\ssk
\ni     
The PCA data showed that the source was extremely variable during our
observation, with many bright flares.
Full details are given in Smith \& Liang (1999).
The bright flares last a few seconds.
Their shape is relatively triangular or has an exponential rise and decay.
Their time profiles are approximately symmetric, usually with slightly
faster decays than rises.
There are many smaller flares that last from tens to hundreds of milliseconds.
We do not see any very short very bright spikes.
There are a few time intervals where the flux rises steadily and then drops
suddenly, sometimes to a level lower than the average before the increase.
Broader brightenings and dimmings that last $\sim 10$ sec are also seen.

The light curves for the 2--5, 5--10, and 10--40 keV bands are almost 
identical, with a tendency for the flares to be slightly softer.
Similarly, hardness ratios between the bands show a weak trend, with the 
source being a little softer at higher fluxes.

The power density spectra (PDS) were also complicated and we
found that broken power laws do not provide adequate fits to any of them.
Instead a pair of zero-centered Lorentzians gives a good general 
description of the shape of the PDS.
We found several quasi-periodic oscillations (QPO), including some that
are harmonically spaced with the most stable frequency at 0.35 Hz.
While the overall rms variability of the source was close to being constant
throughout the observation ($\sim 29$\% integrating between 0.01 and 
50 Hz), there is a small but significant change in the PDS shape with time.
More importantly, we found that {\it the soft 2--5 keV band is more 
variable than the harder 5--10 and 10--40 keV bands, which is unusual for 
this source and for other black hole candidates.}
This result is even more striking bearing in mind that the hard model 
component gives a large if not the dominant contribution to the flux in
the 2--5 keV band, strongly diluting the effect of the soft component.

Cross correlation functions (CCF) show that the light 
curve for the 10--40 keV band lags that of the 2--5 keV band by $\sim 5$ msec,
and it lags that of the 5--10 keV band by $\sim 2.5$ ms.

}

\bsk
\baselineskip = 12pt
{\abstract \ni ACKNOWLEDGMENTS
We thank the referee for some useful clarifications.
This work was supported by NASA grants NAG 5-1547 and 5-3824 at 
Rice University, and NSF grant AST-9417213 at UC Berkeley.
}

\bsk
\baselineskip = 12pt


{\references \ni REFERENCES
\ssk

\ref B\"ottcher, M., Liang, E. P., \& Smith, I. A. 1998, A\&A, 339, 87

\ref Canalizo, G., Koenigsberger, G., Pena, D., \& Ruiz, E. 1995, 
Rev. Mex. Astron. Astrofis., 31, 63

\ref Cowley, A. P., Crampton, D., \& Hutchings, J. B. 1987, AJ, 92, 195

\ref Dove, J. B., et al. 1998, MNRAS, 298, 729

\ref Fender, R. P., Spencer, R. E., Newell, S. J., \& Tzioumis, A. K.
1997, MNRAS, 286, L29

\ref Filippenko, A. V., et al. 1997, PASP, 109, 461

\ref Grabelsky, D. A., et al. 1995, ApJ, 441, 800

\ref Grindlay, J. E. 1979, ApJ, 232, L33

\ref Rubin, B. C., et al. 1998, ApJ, 492, L67

\ref Smith, I. A., \& Liang, E. P. 1999, ApJ, in press (Paper II)

\ref Smith, I. A., Filippenko, A. V., \& Leonard, D. C. 1999, ApJ, in press
(Paper III)

\ref Smith, I. A., et al. 1997,
in Fourth Compton Symposium Part Two (New York: AIP), 932

\ref Smith, I. A., et al. 1999, ApJ, in press (Paper I)

\ref Soria, R., Wickramasinghe, D. T., Hunstead, R. W., \& Wu, K.
1998, ApJ, 495, L95

\ref Sunyaev, R. A., \& Titarchuk, L. G. 1980, A\&A, 86, 121

\ref Tanaka, Y., \& Lewin, W. H. G. 1995, in X-Ray Binaries,
(Cambridge: Cambridge U. Press), 126

\ref Zdziarski, A. A., et al. 1998, MNRAS, in press
(astro-ph/9807300)

} 

\end{document}